\documentclass[letterpaper,usenatbib]{mn2e}

\usepackage{epsfig}
\usepackage{amsmath}
\usepackage{natbib}

\def\apj{ApJ}

\def\apjl{ApJL}
\def\apjs{ApJS}
\def\mnras{MNRAS}
\def\aj{AJ}

\def\nat{Nature}

\newcommand{\hmpc}{h^{-1}{\rm Mpc}}

\newcommand{\zsat}{$z_{\rm sat}$~}
\newcommand{\Nsat}{$N_{\rm sat}$~}

\begin{document}

\title[Testing SHAM in SPH Simulations]{Testing Subhalo Abundance Matching in Cosmological Smoothed Particle Hydrodynamics Simulations}
\author[V. Simha et al]
{Vimal Simha$^{1,2}$, David H. Weinberg$^{1}$, Romeel Dav\'{e}$^{3}$,
\newauthor
Mark Fardal$^{4}$, Neal Katz$^{4}$, Benjamin D. Oppenheimer$^{5}$\\
$^1$ Astronomy Department and Center for Cosmology and AstroParticle
Physics, Ohio State University, Columbus, OH 43210,\\
vsimha,dhw@astronomy.ohio-state.edu\\$^2$ Institute for Computational
Cosmology, Dep.~of Physics, University of Durham, South Road, Durham DH1 3LE\\ 
$^3$University of Arizona,
Steward Observatory, Tuscon, AZ 85721, rad@as.arizona.edu\\
$^4$Astronomy Department, University of Massachusetts at Amherst, MA
01003, fardal,nsk@kaka.astro.umass.edu\\ $^5$Leiden Observatory,
Leiden University, 2300 RA Leiden, The Netherlands;
oppenheimer@strw.leidenuniv.nl\\ }

\maketitle


\begin{abstract}
Subhalo abundance matching (a.k.a. SHAM) is a technique for populating
simulated dark matter distributions with galaxies, assuming a
monotonic relation between a galaxy's stellar mass or luminosity and
the mass of its parent dark matter halo or subhalo. We examine the
accuracy of SHAM in two cosmological smoothed particle hydrodynamics
(SPH) simulations, one of which includes momentum$-$driven galactic
winds. The SPH simulations indeed show a nearly monotonic relation
between stellar mass and halo mass provided that, for satellite
galaxies, we use the mass of the subhalo at the epoch \zsat when it
became a satellite. In each simulation, the median relation for
central and satellite galaxies is nearly identical, though a somewhat
larger fraction of satellites are outliers because of stellar mass
loss. SHAM-assigned masses (at $z=0-2$), luminosities ($R$-band at
$z=0$), or star formation rates (at $z=2$) have a 68\% scatter of
$0.09-0.15$ dex relative to the true simulation values. When we apply
SHAM to the subhalo population of a collisionless N-body simulation
with the same initial conditions as the SPH runs, we find generally
good agreement for the halo occupation distributions and halo radial
profiles of galaxy samples defined by thresholds in stellar
mass. However, because a small fraction of SPH galaxies suffer severe
stellar mass loss after becoming satellites, SHAM slightly
overpopulates high mass halos; this effect is more significant for the
wind simulation, which produces galaxies that are less massive and
more fragile. SHAM recovers the two-point correlation function of the
SPH galaxies in the no-wind simulation to better than 10\% at scales
$0.1\hmpc < r < 10\hmpc$. For the wind simulation, agreement is better
than 15\% at $r > 2\hmpc$, but overpopulation of massive halos
increases the correlation function by a factor $\sim 2.5$ on small
scales. The discrepancy in the wind simulation is greatly reduced if
we raise the stellar mass threshold from 6$\times$10$^9$M$_\odot$ to
3$\times$10$^{10}$M$_\odot$; in this case SHAM overpredicts the SPH
galaxy correlation function by $\sim$20\% at $r < 1\hmpc$ but agrees
well with SPH clustering at larger scales.

\end{abstract}


\newpage
\section{Introduction}

In the standard theoretical description of galaxy formation, galaxies
form by the dissipation of the baryonic component within collisionless
dark matter halos \citep[e.g.][]{white78,fall80}. When a dark matter
halo enters the virial radius of a more massive halo, it is subjected
to tidal stripping and potentially to disruption. Nonetheless, high
resolution N-body simulations have indicated that massive halos retain
a substantial amount of
substructure \citep{klypin99,moore99,springel01}, consisting of bound
dark matter clumps orbiting within the potential of their host
halo. Evidently, such subhalos were themselves independent,
self-contained halos in the past, before merging with a more massive
halo. If sufficiently massive, these subhalos were sites of baryon
dissipation and star formation in the past. There are many indications
from studies of the statistical properties of how galaxies and
substructures populate halos that galaxies in groups and clusters are
in fact the observational counterparts of subhalos. For
example, \cite{colin99} and \cite{kravtsov04} show that the
correlation functions of substructures in high resoution N-body
simulations are in good agreement with the observed correlation
functions of galaxies. On the theory side, \cite{kravtsov04} find that
the distribution of subhalos in high resolution N-body simulations is
similar to that of smoothed particle hydrodynamics (SPH) galaxies
in \cite{berlind03} and \cite{zheng05}.

Subhalo abundance matching (SHAM) is a technique for assigning
observable galaxy properties to a halo/subhalo population in an N-body
simulation. It is based on assuming a monotonic relationship between
observable properties of galaxies and dynamical properties of dark
matter substructures. As subhalos that fall into the virial radius of
more massive halos are subjected to stripping and tidal disruption,
several authors \citep[e.g.][]{conroy06,vale06,moster10} contend that
the properties of satellite galaxies should be better correlated with
the properties of subhalos at the time of their accretion onto a more
massive halo rather than their present day properties. By using the
SHAM technique, with this accretion-epoch matching, \cite{conroy06}
match the observed luminosity dependence of galaxy clustering at a
wide range of epochs, ranging from $z=0$ to $z\sim5$. Assuming a
monotonic relationship between galaxy mass and halo mass and using the
same SHAM technique, \cite{guo10} reproduce the observationally
inferred relation between stellar mass and halo virial
mass. \cite{gomez10} use the abundance matching technique to match the
observed relations between stellar mass and circular velocity and
luminosity and circular velocity, and to match the estimated galaxy
velocity function.
 
In this paper, we investigate the effectiveness of subhalo abundance
matching (SHAM) in cosmological SPH simulations, where we know exactly
the relation between the properties of galaxies and the masses of
their parent halos and subhalos. We investigate the degree to which
there is a direct correspondence between the properties of dark matter
substructures in a dissipationless numerical simulation of a
cosmological volume and condensed baryons in numerical simulations
(with the same initial conditions) that include a dissipative
component. We extend the similar study of \cite{weinberg08} in several
ways. Firstly, our simulation volume is more than ten times
larger. Secondly, one of our dissipative simulations includes ejective
feedback in the form of momentum driven winds, which curtail star
formation and produce a stellar mass function that is in better
agreement with the observations. Lastly and perhaps most importantly,
we relate galaxy properties to the subhalo mass at the epoch of
accretion rather than at the present time (see comparison in \S4). We
make a direct assessment of the effectiveness of SHAM as a method for
assigning stellar mass, luminosity or star formation rate to subhalos
and investigate the sources of its breakdown.

Our investigation offers insight into the physical mechanisms that
shape galaxy masses and luminosities in these simulations, and it also
has practical import. If the subhalo population in the purely
gravitational simulation does indeed trace the observable properties
of the galaxy population, it enables us to make observable predictions
of quantities like the luminosity dependence of galaxy clustering
based on computationally less expensive N-body simulations instead of
hydrodynamic simulations. SHAM also offers a relatively inexpensive
tool for creating artificial galaxy catalogues to support statistical
analyses of large scale structure data sets.

In \S2, we describe our simulation, our methods for identifying halos,
subhalos and galaxies, and the subhalo abundance matching
scheme. In \S3, we investigate the relationship between the stellar
masses, luminosities, and star formation rates of SPH galaxies and
dark matter substructures in our SPH simulations to test some of the
underlying assumptions of the subhalo abundance matching
technique. In \S4, we investigate whether the SPH galaxy population
can be recovered from the subhalo population in our matched N-body
simulation; note that it is this comparison rather than the
investigations in \S3 that tests SHAM as it has been traditionally
implemented. Finally, in \S5, we summarise our results and discuss
their implications.

\section{Methods}

\subsection{Simulations}

Our simulations are performed using the GADGET-2
code \citep{springel05} as modified
by \cite{oppenheimer08}. Gravitational forces are calculated using a
combination of the Particle Mesh algorithm \citep{hockney81} for large
distances and the hierarchical tree
algorithm \citep{barnes86,hernquist87} for short distances. The
smoothed particle hydrodynamics (SPH) algorithm is entropy and energy
conserving and is based on \cite{springel02}.  The details of the
treatment of radiative cooling can be found in \cite{katz96}. Gas can
dissipate energy via Compton cooling and radiative cooling, computed
assuming a primordial gas composition and a background UV flux based
on \cite{haardt01}. The details of the treatment of star formation can
be found in \cite{springel03}. Briefly, each gas particle satisfying a
temperature and density criterion is assigned a star formation rate,
but the conversion of gaseous material to stellar material proceeds
stochastically. The parameters for the star formation model are
selected so as to match the $z=0$ relation between star formation rate
and gas density \citep{kennicutt98,schmidt59}.

We adopt a $\Lambda$CDM cosmology (inflationary cold dark matter with
a cosmological constant) with $\Omega_m=0.25$,
$\Omega_{\Lambda}=0.75$, $h\equiv H_0/100$ km s$^{-1}$Mpc$^{-1}$=0.7,
$\Omega_b=0.044$, spectral index $n_s=0.95$, and the amplitude of the
mass fluctuations scaled to $\sigma_8=0.8$. These values are
reasonably close to current estimates from the cosmic microwave
background \citep{larson10} and large scale
structure \citep{reid10}. We do not expect minor changes in the values
of the cosmological parameters to affect our conclusions.

We follow the evolution of $288^3$ dark-matter particles and $288^3$
gas particles, i.e. just under 50 million particles in total, in a
comoving box that is 50$h^{-1}$ Mpc on each side, from $z=129$ to
$z=0$. The dark matter particle mass is 4.3 $\times$ $10^8$
$M_{\odot}$, and the SPH particle mass is 9.1 $\times$ $10^7$
$M_{\odot}$. The gravitational force softening is a comoving 5$h^{-1}$
kpc cubic spline, which is roughly equivalent to a Plummer force
softening of 3.5$h^{-1}$ kpc.

One of our simulations, SPHw (SPH winds) incorporates kinetic feedback
through momentum driven winds as implemented by \cite{oppenheimer06}
and \cite{oppenheimer08} where the details of the implementation can
be found. Briefly, wind velocity is proportional to the velocity
dispersion of the galactic halo, and the ratio of the gas ejection
rate to the star formation rate is inversely proportional to the
velocity dispersion of the galactic halo.

We also carry out a simulation with the same cosmological and
numerical parameters as the SPHw simulation, but without momentum
driven winds, SPHnw (SPH No Winds). Although no energy is kinetically
imparted to SPH particles, the SPHnw simulation includes thermal
feedback from supernovae \citep[see][]{springel03}. However, since the
surrounding gas is dense, the energy is radiated away before it can
drive outflows or significantly suppress star formation. Our SPHnw and
SPHw simulations are analogous to the nw and vzw simulations
of \cite{oppenheimer10}, who investigate the growth of galaxies by
accretion and wind recycling and compare predicted mass functions to
observations. Our simulations are also used by \cite{zu10}, who
investigate intergalactic dust extinction.

In addition to the two SPH simulations, we carry out a
non-dissipative, purely gravitational, N-body simulation with
identical cosmological and numerical parameters and the same initial
positions and velocities of particles as the SPH simulations, except
that the dark matter particle mass is higher by a factor of
$\Omega_m/(\Omega_m-\Omega_b)$ to compensate for the gravitational
effects of not including baryons. We refer to this N-body simulation
as the DM (dark matter only) simulation.

\subsection{Identification of Groups and Substructures}

We identify dark matter haloes using a FOF (friends-of-friends)
algorithm
\citep{davis85}. The algorithm selects groups of particles in which
each particle has at least one neighbour within a linking length. We
assign a mass to the halos using a SO (spherical overdensity)
criterion, with the threshold density set to the virial overdensity in
spherical collapse \citep{kit96}. At $z=0$, the mean interior
overdensity with respect to the critical density of the Universe is
94. We set the centre of the group at the most bound FOF particle and
go out in radius until the mean density enclosed is equal to the
virial density. While FOF occasionally links multiple distinct
concentrations together, most FOF halos contain a distinct mass
concentration (see Figures 1 and 2 of \cite{simha09}).

To identify substructures within halos, we use the
AdaptaHOP \citep{aubert04} code. The details of the algorithm can be
found in
\cite{aubert04}, so we provide only a brief summary of it here. We calculate densities around each dark matter 
particle using an SPH-like kernel estimator, with a cubic spline
kernel containing 32 neighbours. We then partition the ensemble of
particles into ``peak patches'' where a peak patch is a set of
particles with the same local density maximum. The connectivity
between peak patches is dictated by the saddle points in the density
field. We identify subsets of particles in each peak patch with SPH
density larger than the density of the highest saddle point connecting
it to a neighbouring peak patch. To select statistically significant
substructures as opposed to random fluctuations in the density field,
we require that the mean density of a substructure be 4$\sigma$ above
the SPH density of the saddle point. The performance of AdaptaHOP on a
simulation similar to that here is illustrated in \cite{weinberg08}
(see their Figure 1).

Hydrodynamic cosmological simulations that incorporate cooling and
star formation produce dense groups of baryons with sizes and masses
comparable to the luminous regions of observed
galaxies \citep{katz92,evrard94}. We identify galaxies using the
Spline Kernel Interpolative DENMAX
(SKID\footnote{http://www-hpcc.astro.washington.edu/tools/skid.html})
algorithm \citep{gelb94,katz96}, which identifies gravitationally
bound particles associated with a common density maximum. We refer to
the groups of stars and cold gas thus identified as galaxies. The
simulated galaxy population becomes substantially incomplete below our
resolution threshold of $\sim$64 SPH particles \citep{murali02}, which
corresponds to a baryonic mass of 5.8 $\times 10^9$
$M_{\odot}$. Although this threshold applies to the total baryonic
mass (stars plus cold, dense gas) of galaxies, we adopt it as our
threshold for stellar mass and ignore galaxies with lower stellar
mass.

\begin{figure}
\centerline{
\epsfxsize=4.5truein
\epsfbox{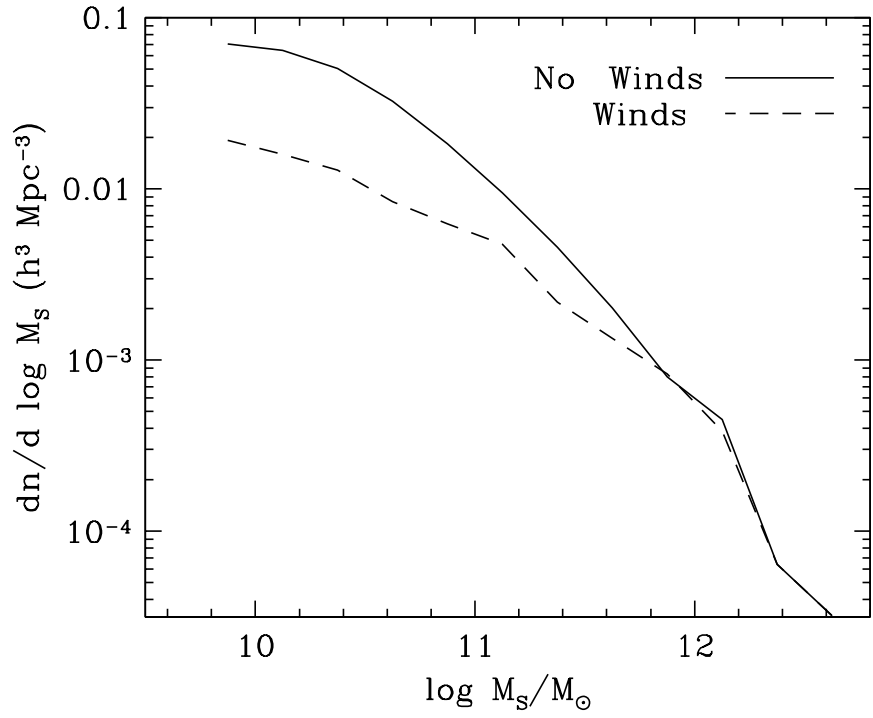}
}
\caption{
Galaxy stellar mass functions at $z=0$ in the SPHw simulation
(dashed), which incorporates momentum driven winds, and the SPHnw
simulation (solid), which does not. In this and later plots, M$_S$
refers to the stellar mass of SKID-identified galaxies.  }
\label{fig:sub0}
\end{figure}

Figure \ref{fig:sub0} compares the galaxy stellar mass function in the
SPHnw and SPHw simulations at $z=0$.  In the SPHnw simulation, there
are 7,952 galaxies above our resolution threshold, corresponding to a
space density of 0.064 $h^3$ Mpc$^{-3}$. In the SPHw simulation there
are only 2,264 galaxies above our resolution threshold, corresponding
to a space density of 0.018 $h^3$ Mpc$^{-3}$ because wind feedback
pushes the stellar mass of many galaxies below the 64 m$_{\rm SPH}$
threshold. The two mass functions gradually converge towards higher
masses, joining at $M_S$ $>$ 10$^{11.8}$ M$_{\odot}$, because wind
feedback has less suppressing effect in larger systems since the
amount of material ejected in this model scales inversely with
circular velocity. \cite{oppenheimer10} discuss the comparison between
the predicted stellar mass functions and observational estimates in
some detail. Roughly speaking, the SPHw model reproduces observational
estimates for $M_S$ $<$ 10$^{11}$ M$_{\odot}$ , but it predicts
excessive galaxy masses (at a given space density) for $M_g$ $>$
10$^{11.8}$ M$_{\odot}$.

\subsection{Subhalo Abundance Matching}

Subhalo abundance matching (SHAM) is a technique for assigning
galaxies to simulated dark matter halos and subhalos. The essential
assumptions are that all galaxies reside in identifiable dark matter
substructures and that luminosity or stellar mass of a galaxy is
monotonically related to the potential well depth of its host halo or
subhalo. Some implementations use the maximum of the circular velocity
profile as the indicator of potential well depth, while others use
halo or subhalo mass. The first clear formulations of SHAM as a
systematic method appear in \cite{conroy06} and \cite{vale06}, but
these build on a number of previous studies that either test the
underpinnings of SHAM or implicitly assume SHAM-like galaxy
assignment \citep[e.g.][]{colin99,kravtsov04,nagai05}.

N-body simulations produce subhalos that are located within the virial
radius of SO halos. The present mass of subhalos is a product of mass
build up during the period when the halo evolves in isolation and
tidal mass loss after it enters the virial radius of a more massive
halo \citep[e.g.][]{kravtsov04,kazantzidis04}. The stellar component,
however, is at the bottom of the potential well and more tightly bound
making it less likely to be affected by tidal forces. Therefore,
several authors \citep[e.g.][]{conroy06,vale06} argue that the
properties of the stellar component should be more strongly correlated
with the subhalo mass at the epoch of accretion rather than at $z=0$.

\cite{vale06} apply a global statistical correction to subhalo masses relative to halo masses (as do \citealt{weinberg08}), while \cite{conroy06} explicitly identify subhalos at the epoch of accretion and use the maximum circular velocity at that epoch. Our formulation here is similar to that of \cite{conroy06}, though we use mass rather than circular velocity. Specifically, we assume a monotonic relationship between stellar mass and halo mass and determine the form of this relation by solving the implicit equation
\begin{equation}
 n_S(>M_S) = n_H(>M_H),
\end{equation}
where $n_S$ and $n_H$ are the number densities of galaxies and halos,
respectively, $M_S$ is the galaxy stellar mass threshold, and $M_H$ is
the halo mass threshold chosen so that the number density of halos
above it is equal to the number density of galaxies in the sample.
The quantity $M_{\rm H}$ is defined as follows:
\begin{equation}
M_H=\begin{cases} M_{\rm halo} (z=0) & \text{for distinct halos},\\
M_{\rm halo} (z=z_{\rm sat})& \text{for subhalos},
\end{cases}
\end{equation}
where z$_{\rm sat}$ is the epoch when a halo first enters the virial
radius of a more massive halo. We also consider (in \S3.2) variants of
this procedure in which we substitute R-band luminosity or
instantaneous star formation rate for stellar mass in eq.(1).

In \S3, we use our SPH simulations to test the degree to which galaxy
stellar mass and luminosity are monotonic functions of halo/subhalo
mass, as assumed in SHAM. For independent halos, we set $M_H$ equal to
the $z=0$ mass of SO halos. We use the AdaptaHOP code to identify
subhalos hosting satellite galaxies above the resolution threshold at
$z=0$ in our SPH simulations. We then use their particle membership to
identify their progenitor SO halos at $z_{\rm sat}$, whose mass we
adopt as $M_H$. We can identify a host subhalo for the vast majority
of our galaxies at $z=0$. However, for about one percent of galaxies
we cannot identify any associated substructure. In these cases, we use
the galaxy's baryonic particles to trace its high redshift progenitors
up to the epoch when it was the only galaxy in an SO halo and adopt
the mass of this SO halo as $M_H$. After matching, we examine the
correlations of galaxy properties with $M_H$ for both central and
satellite systems, and we compare the stellar masses, luminosities,
and (at high redshift) star formation rates that would be assigned by
monotonic matching to the simulation values. Of course, SHAM is
usually applied to collisionless N-body simulations, not SPH
simulations, and the subhalo populations can differ even for the same
initial conditions because of the dynamical effects of the dissipative
baryons on the dark matter.

In \S4, we populate halos/subhalos in our DM-only simulation using
SHAM, following a procedure akin to the one usually used to match
simulated halos/subhalos to observed galaxies. For independent halos,
we set $M_H$ equal to the $z=0$ mass of SO halos. We identify subhalos
using the AdaptaHOP code and use their particle membership to identify
their progenitor SO halos at $z_{\rm sat}$, whose mass we adopt as
$M_H$. We have checked that at \zsat, the relation between $v_{\rm
max}$ and $M_H$ is close to monotonic, so we would get similar results
from using $v_{\rm max}$ instead of $M_H$. For comparison, we also
implement a procedure that we refer to as SHAMz0, where we set $M_H$
equal to the $z=0$ mass for both independent halos and subhalos
identified in our N-body simulation.

\cite{knebe11} find differences in the performance of various halo finding algorithms in identifying substructures and accurately recovering their properties, particularly for subhalos containing fewer than 40 particles. Since we trace the \zsat progenitors of $z=0$ substructures, our results are unlikely to be affected by random fluctuations in the density field of halos that may be spuriously identified as substructures. However, if subhalos hosting satellite galaxies that have merged with the central galaxy of the halo are identified as substructures, we would overestimate the halo occupation of massive halos. Conversely, subhalos that fall into more massive halos and lose a substantial fraction of their mass due to tidal stripping may no longer be resolved in the simulation at $z=0$ even though they might survive and host satellite galaxies in a sufficiently high resolution calculation.

\section{SHAM in the SPH Simulations}

\subsection{Stellar Mass}

\begin{figure}
\centerline{
\epsfxsize=5.5truein
\epsfbox{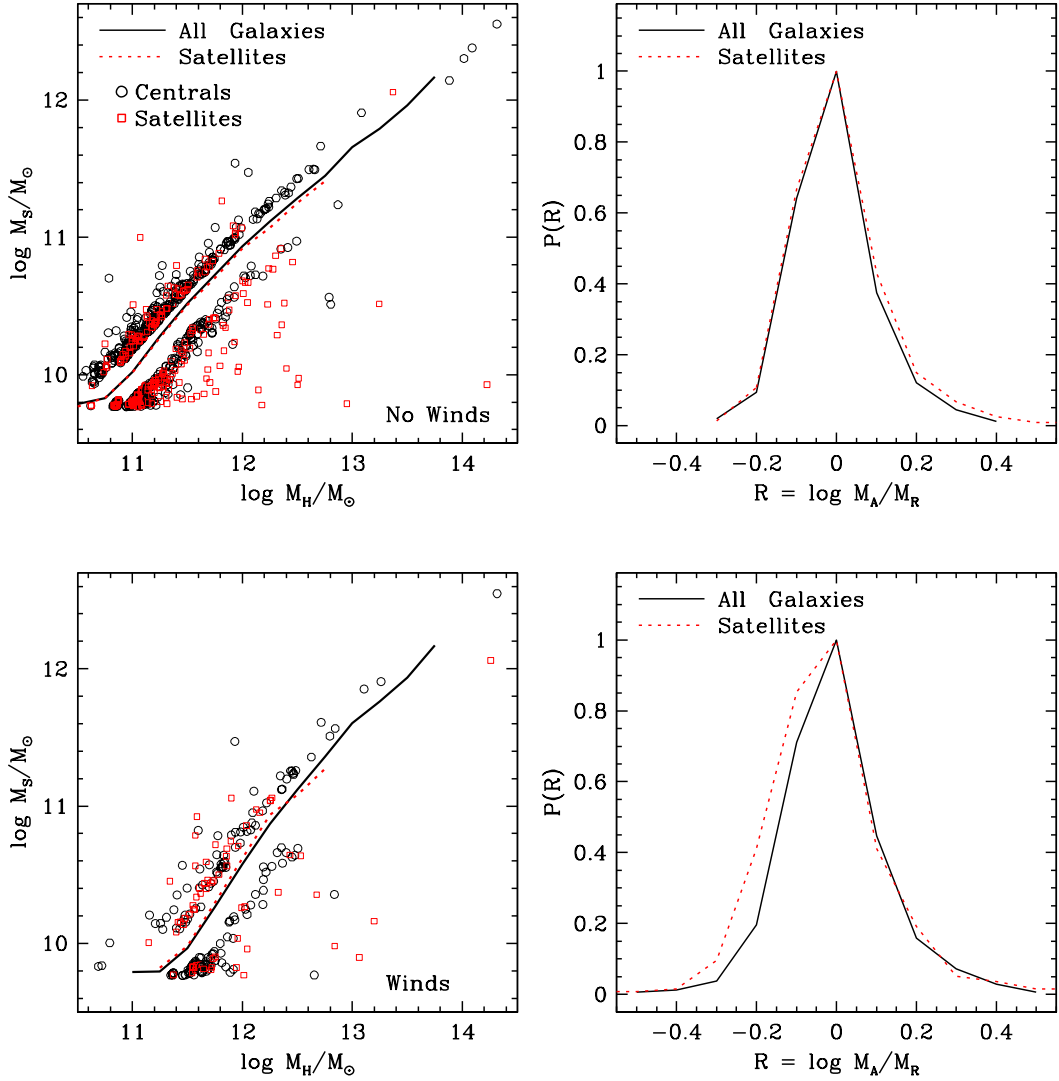}
}
\caption{
(Left) Stellar mass at $z=0$ versus halo mass in the SPHnw simulation
(top) and the SPHw simulation (bottom). Each point represents an SPH
galaxy and we only include galaxies above the $M_S$ = 64m$_{\rm SPH}$
$=$ 5.8 $\times$ 10$^9$ M$_{\odot}$ threshold at $z=0$. For central
galaxies, shown as black points, the halo mass is the $z=0$ mass of
the host halo, while for satellite galaxies, shown as red points, the
halo mass is the mass of the parent halo just before \zsat, the epoch
at which it became a satellite. Only galaxies that are in the top 5\%
or bottom 5\% by stellar mass in each 0.25 decade wide halo mass bin
(relative to all galaxies in the bin) are shown. The solid and dotted
curves show the median stellar mass in each halo mass bin for all
galaxies and satellite galaxies respectively. (Right) Probability
distribution of the ratio of stellar mass assigned by subhalo
abundance matching to SPH galaxy mass in the SPHnw simulation (top)
and SPHw simulation (bottom), with the solid curve standing for all
galaxies and the dotted curve for satellite galaxies.  }
\label{fig:sub1}
\end{figure}

The left column of Figure \ref{fig:sub1} shows galaxy stellar mass
plotted against halo mass in the SPHnw simulation (top) and the SPHw
simulation (bottom). For galaxies that are the central objects of
their parent halos (henceforth central galaxies), the halo mass,
$M_H$, on the horizontal axis, is the mass of the SO halo in which the
galaxy is located at $z=0$. For galaxies that are not the central
objects of their respective SO halos (henceforth satellite galaxies),
$M_H$ is the mass of the SO halo in which the galaxy is located at
$z_{\rm sat}$, the last output epoch before its parent halo fell into
the virial radius of a more massive halo; $z_{\rm sat}$ values range
between $z=3$ and $z=0.05$. The black solid curve and red dotted curve
show the median galaxy stellar mass in evenly spaced logarithmic bins
of halo mass for all galaxies and satellite galaxies respectively,
while the red open circles and black open rectangles show individual
galaxies that are within the top five percent and bottom five percent
by stellar mass in each halo mass bin.

The key result of Figure \ref{fig:sub1} is that the ratio of galaxy
mass to $M_H$ is similar for central galaxies and satellite
galaxies. Between $z_{\rm sat}$ and $z=0$, satellite galaxies have
lower growth rates compared to central galaxies of similar
mass \citep[see][]{keres09,simha09}. However, during the same time
period, while central galaxies grow at a faster rate, their host halos
also accrete mass and ``receive'' mergers of lower mass halos. The
balance between stellar mass growth and halo mass growth leads to a
similar $M_S$-$M_H$ relation at $z=0$ for central and satellite
systems, in both simulations.

Note that in Figure \ref{fig:sub1} as well as in the remainder of this
section, we only consider galaxies that are above our resolution
threshold at $z=0$, with $M_S$ $>$ 64$m_{\rm SPH}$ = 5.8$\times$10$^9$
M$_{\odot}$. However, some satellite galaxies in dense environments
experience mass loss between $z_{\rm sat}$ and $z=0$, and,
consequently, a fraction of satellites that are above the resolution
threshold at $z_{\rm sat}$ are pushed below it by $z=0$. We defer
examination of this point to \S4.

We assign galaxies to the halo/subhalo population via the monotonic
mapping procedure described in \S2. The right-hand side column of
Figure \ref{fig:sub1} shows the distribution of the ratio of the
stellar mass assigned to a halo to the stellar mass of the SPH galaxy
that is actually located within it for the SPHnw simulation (top) and
the SPHw simulation (bottom). In addition to the distribution for all
galaxies (black solid curve), we show the distribution for satellite
galaxies alone (red dotted curve). For most halos, the SHAM assigned
mass is close to the mass of the SPH galaxy located within it in both
the SPHnw and the SPHw simulations, although there are a small number
of extreme outliers. We characterise the width of the distributions by
$\sigma_M$ such that 68\% of galaxies have $|R|$ = $|$log $M_A$/$M_R|$
$<$ $\sigma_M$ where $M_A$ is the assigned mass and $M_R$ is the SPH
(real) mass. In the SPHnw simulation, the width of the distribution is
$\sigma_M$ = 0.09, while in the SPHw simulation $\sigma_M$ = 0.11. In
both simulations, the distribution of SHAM assigned masses for
satellite galaxies is also centered on the SPH mass, but there is
greater scatter than in the case of central galaxies. For satellite
galaxies in the SPHnw simulation, the distribution of assigned mass is
skewed such that the assigned mass is likely to be slightly higher
than the SPH mass.

\begin{figure}
\centerline{
\epsfxsize=5.5truein
\epsfbox{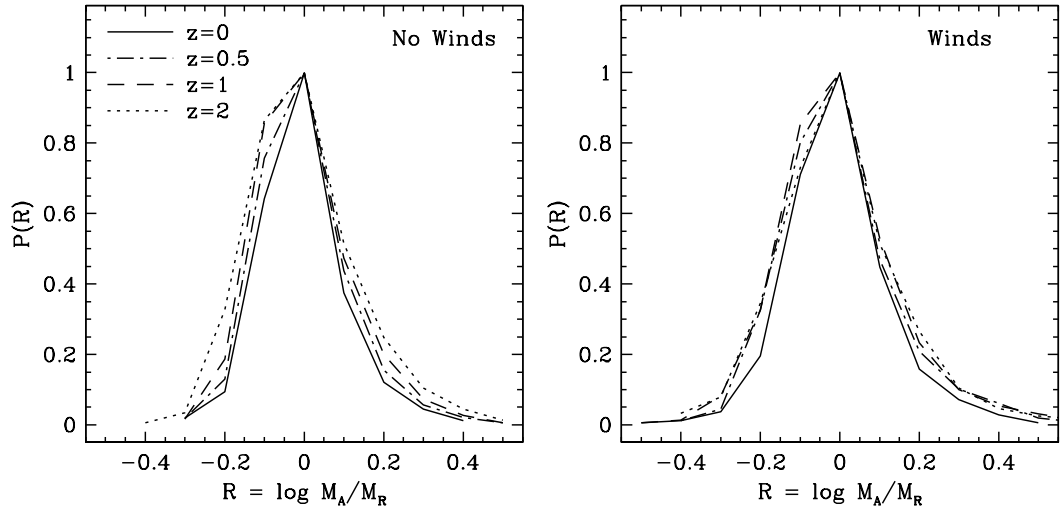}
}
\caption{
Probability distribution of the ratio of stellar mass assigned by
subhalo abundance matching to SPH galaxy mass in the SPHnw simulation
(left) and SPHw simulation (right). The solid, dot-dashed, dashed and
dotted curves represent redshifts, $z=0$, 0.5, 1 and 2 respectively.
}
\label{fig:sub2}
\end{figure}

Figure \ref{fig:sub2} presents the same distribution of $|$log
$M_A$/$M_R|$ at $z = 0.5, 1$ and 2, and compared to the $z=0$ result
from Figure \ref{fig:sub1}. There is good agreement between the SHAM
assigned stellar masses and the SPH stellar masses at all four
epochs. The fraction of satellite galaxies decreases with increasing
redshift. Despite this, $\sigma_M$ shows a continuous trend of
increasing with redshift, from 0.09 at $z=0$ to 0.13 at $z=2$ in the
SPHnw simulation, and from 0.11 at $z=0$ to 0.14 at $z=2$ in the SPHw
simulation. Although the total baryonic mass (not shown) is equally
well correlated with halo mass at $z=2$ and at $z=0$, the higher mean
gas fraction and the larger halo-to-halo scatter in gas fraction at
high redshift leads to higher scatter in the halo mass-stellar mass
relation.

\begin{figure}
\centerline{
\epsfxsize=7.5truein
\epsfbox{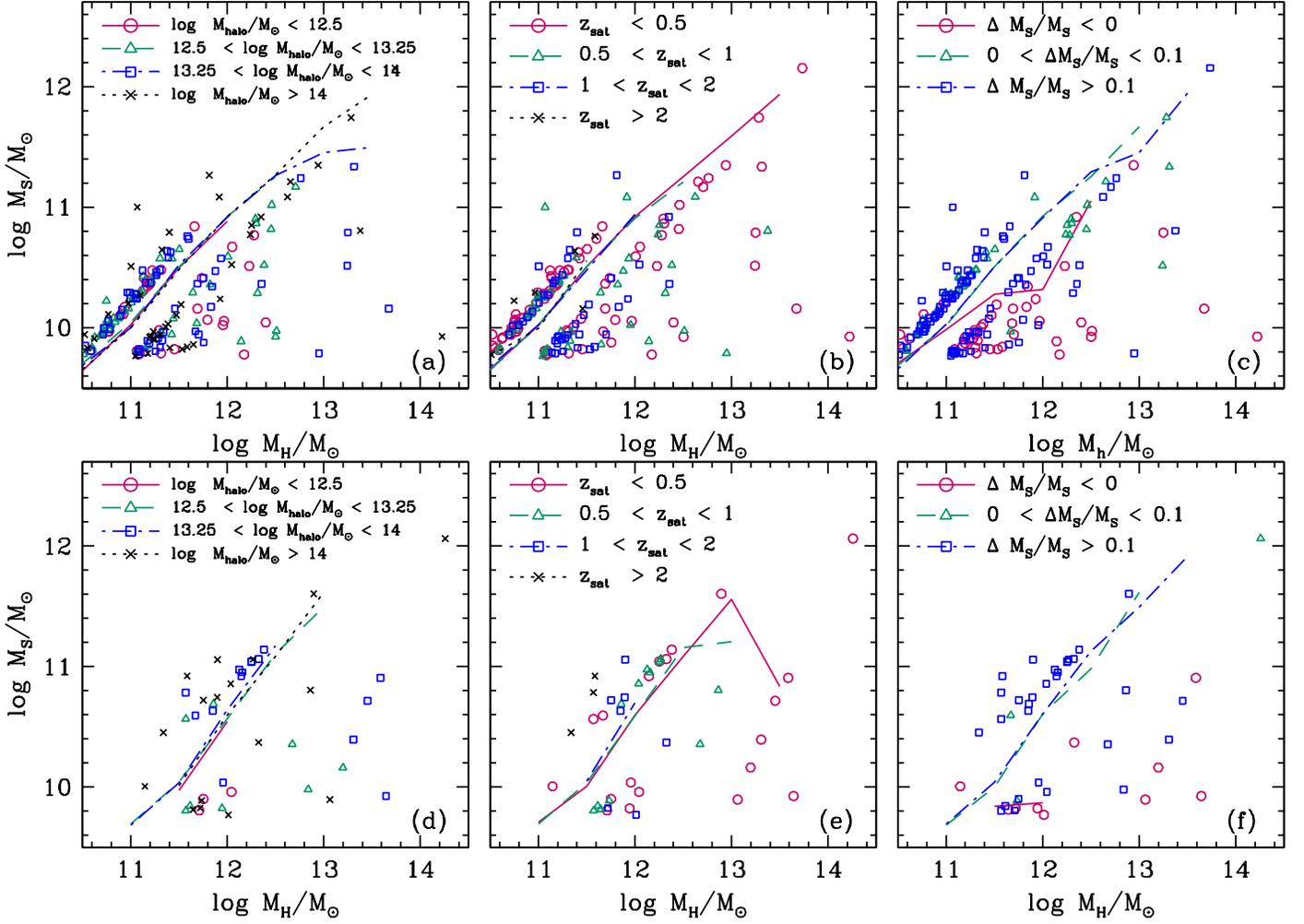}
}
\caption{
Stellar mass of satellite galaxies versus parent halo mass just before
$z_{\rm sat}$. Each point represents a galaxy, and only galaxies that
are in the top 5\% or bottom 5\% by stellar mass in each 0.25 decade
wide halo mass bin (relative to the distribution of all galaxies) are
shown. In panel (a), the circles, triangles, squares and crosses stand
for different $z=0$ host halo mass bins while in panels (b) and (c),
they stand for bins of $z_{\rm sat}$, the epoch of accretion of the
satellite and $\Delta M / M$, the change in stellar mass since $z_{\rm
sat}$, respectively. Panels (d), (e) and (f) are analogous to panels
(a), (b) and (c) but using the SPHw simulation.  }
\label{fig:sub3}
\end{figure}

While the relationship between halo mass and galaxy stellar mass is
roughly monotonic in our SPH simulations, there is some scatter, with
the strongest outliers arising from satellite galaxies. We examine the
sources of this scatter in Figure \ref{fig:sub3}, both to understand
the physical processes that give rise to it and to explore the
possibility of adding a parameter that would sharpen the subhalo
abundance matching. We restrict ourselves to $z=0$ as the satellite
galaxy sample is largest at this epoch. Each panel plots satellite
galaxy stellar mass as a function of $M_H$ at $z_{\rm sat}$ with SPHnw
in the top panels and SPHw in the bottom.

In panel (a), points are colour coded by the mass of the $z=0$ halo
(not subhalo) that hosts the satellite, and lines show the median
relation between $M_S$($z=0$) and $M_H$(\zsat) in four bins of $z=0$
halo mass. The median curves are nearly identical for the four bins,
indicating that the typical $M_S$/$M_H$ is at most minimally
correlated with the final halo mass. The outlier points at low
$M_S$/$M_H$ are mostly galaxies that have experienced stellar mass
loss. These outliers are found in all $M_{\rm halo}$ bins; one should
not read too much into the relative numbers of outlier points as the
total number of satellites varies from bin to bin. We caution that
this figure does not show galaxies that have lost enough mass to fall
below the $M_S$ = 64 $m_{\rm SPH}$ threshold by $z=0$ (see Figure 11,
below).

In panel (b), points and lines are coded by satellite accretion
epoch. Once again, there is little difference in the median relations
and outliers are found in all the \zsat bins (except \zsat $>$
2). Panel (c) divides galaxies into those that have lost stellar mass
since $z_{\rm sat}$, those that have increased their stellar mass by
less than 10\% since \zsat and those that have increased their stellar
mass by more than 10\%. Not surprisingly, galaxies that have lost
stellar mass have systematically lower $M_S$/$M_H$ at $z=0$. Median
relations for the other two populations are similar. There are some
galaxies that are low $M_S$/$M_H$ outliers despite having gained mass
since $z_{\rm sat}$, indicating that at least some of this outlier
population comes from satellites that had anomalously low $M_S$ at
$z_{\rm sat}$. Panels (d) - (f) repeat this analysis for the SPHw
simulation. The results are qualitiatively similar, though the smaller
number of galaxies in this simulation makes it difficult to draw firm
conclusions.

\subsection{Luminosity and Star Formation Rate}

So far, we have used SHAM to assign stellar masses to halos and
compare them to the stellar masses of SPH galaxies located in those
halos. However, when SHAM is applied to an observed galaxy population,
it is often used to assign luminosities to simulated halos to compare
the luminosity dependence of the clustering properties of the
simulated halos to that of the observed galaxies.

The luminosity of a galaxy is correlated with its stellar mass, but is
not reducible to a simple function of stellar mass because the stars
formed at different times. We track the star formation histories of
simulated galaxies in our SPH simulations, and use the stellar
population synthesis package of \cite{conroy09} to compute their
luminosities. We assume that stars are formed with a Chabrier initial
mass function. We assume solar metallicity and do not consider dust
extinction. Changing the initial mass function in the same way for all
galaxies would alter the zero-point of the luminosity-stellar mass
relation but would not be likely to add scatter, while including dust
extinction would shift the mean relation and increase the scatter
somewhat.

\begin{figure}
\centerline{
\epsfxsize=5.5truein
\epsfbox{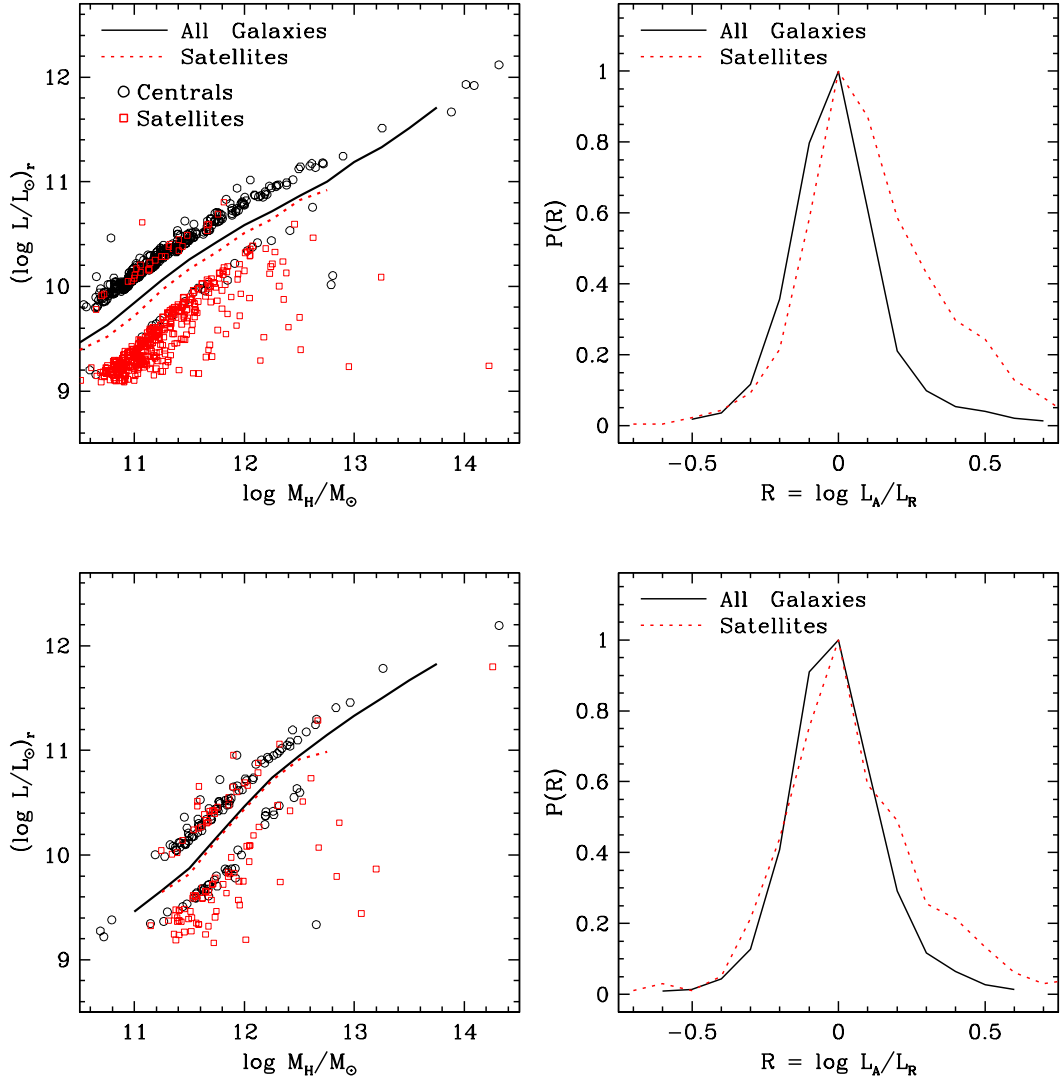}
}
\caption{
(Left) $R$-band luminosity at $z=0$ versus halo mass in the SPHnw
simulation (top) and the SPHw simulation (bottom). Each point
represents a galaxy in the SPHnw simulation. For central galaxies,
shown as black points, the halo mass is the $z=0$ mass of the host
halo, while for satellite galaxies, shown as red points, the halo mass
is the mass of the parent halo just before \zsat. Only galaxies that
are in the top 5\% or bottom 5\% by $R$-band luminosity in each 0.25
decade wide halo mass bin are shown. The solid and dotted curves show
the median $R$-band luminosity in each halo mass bin for all galaxies
and satellite galaxies respectively. (Right) Probability distribution
of the ratio of $R$-band luminosity assigned by subhalo abundance
matching to $R$-band luminosity of SPH galaxies computed using a
stellar population synthesis code in the SPHnw simulation (top) and
SPHw simulation (bottom), with the solid curve standing for all
galaxies and the dotted curve for satellite galaxies.  }
\label{fig:sub4}
\end{figure}

The left column of Figure \ref{fig:sub4} shows the $r$-band luminosity
of SPH galaxies against halo mass in the SPHnw simulation (top) and
the SPHw simulation (bottom). For galaxies that are the central
objects of their parent halos (henceforth central galaxies), the halo
mass, $M_H$, on the horizontal axis, is the mass of the SO halo in
which the galaxy is located at $z=0$. For satellite galaxies, it is
the mass of the SO halo in which the galaxy is located at \zsat, the
last output epoch before its parent halo fell into the virial radius
of a more massive halo. The points show individual galaxies that are
either in the top 5\% or bottom 5\% by luminosity in each halo mass
bin. Most of the low luminosity outliers are satellite galaxies. This
is primarily because satellite galaxies typically have an older
stellar population than central galaxies of similar stellar mass and
are, therefore, less luminous. A secondary factor, of less importance,
is the difference between central and satellite galaxies in the
stellar mass-halo mass relation shown in Figure \ref{fig:sub1}.

To assign luminosities to our halos, we rank order halos by mass and
assign simulated galaxies rank ordered by $r$-band luminosity to them,
as done previously for stellar mass. The right column of
Figure \ref{fig:sub4} shows the distribution of the ratio of the SHAM
assigned luminosity of a halo to the luminosity of the SPH galaxy
within it in the SPHnw simulation (top right) and the SPHw simulation
(bottom right). In both simulations, the distribution of SHAM assigned
luminosities is centered on the SPH luminosity. In analogy with the
previous subsection, we define $\sigma_L$ such that 68\% of galaxies
have $|R|$ = $|$log L$_A$/L$_R|$ $<$ $\sigma_L$ where L$_A$ is the
assigned luminosity and L$_R$ is the SPH (real) luminosity. In the
SPHnw as well as the SPHw simulation, $\sigma_L$ = 0.15, which is
greater than the corresponding $\sigma_M$ (see
Figure \ref{fig:sub1}). For satellite galaxies, the assigned
luminosity is systematically higher than the SPH luminosity because of
the stellar population differences, the offset being 0.08 dex in the
SPHnw simulation but only 0.02 dex in the SPHw simulation.

The clustering properties of $z$ $\geq$ 2 galaxies are studied
observationally using the Lyman break technique, in which high
redshift star forming galaxies are identified by optical photometry
alone using their redshifted rest frame UV
radiation \citep{steidel96,steidel99,steidel03}. \cite{conroy06} find
good agreement between the clustering of Lyman break galaxies (LBGs)
and the clustering of halos in their N-body simulation when they use
SHAM to match halos to galaxies. The relationship between LBGs and
their host halos is important in understanding the properties of LBGs,
in particular whether they are a quiescent star-forming
population \citep{coles98,mo99,giavalisco01} or a merger driven
starbust population \citep{sommerville01,scan03}.
These applications raise the question of whether SHAM can
be reliably applied to the rest-frame UV luminosities
of high redshift galaxies, which depend mainly on their
star formation rates (SFRs) rather than their stellar masses.

\begin{figure}
\centerline{
\epsfxsize=5.5truein
\epsfbox{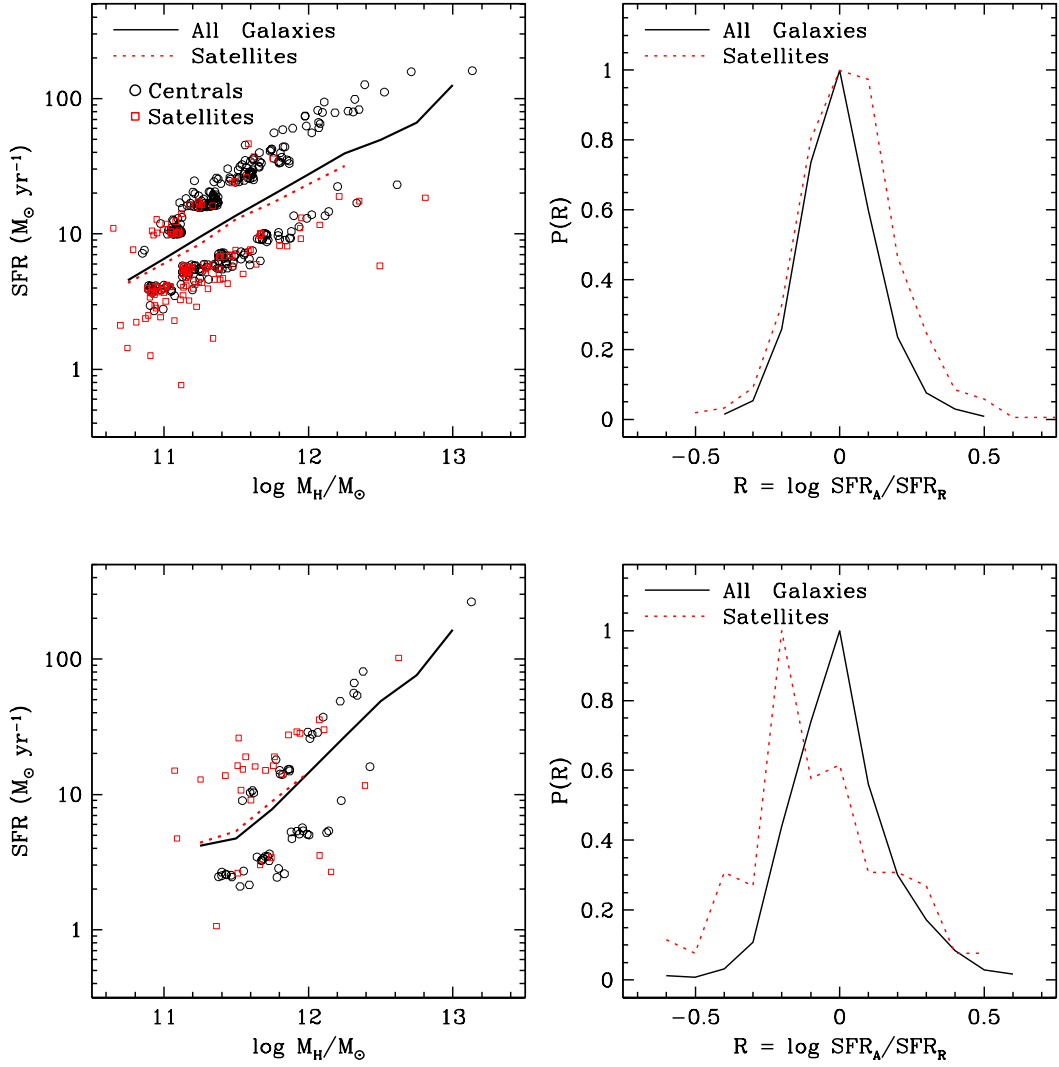}
}
\caption{
Relation between galaxy SFR and stellar mass (left), and accuracy of
SHAM SFR assignment (right), for galaxies in the SPHnw (top) and SPHw
(bottom) simulations at $z=2$. The format is the same as Figures 2 and
5, but with instantaneous SFR used in place of stellar mass or
$R$-band luminosity. Only galaxies with stellar mass above the 64
$m_{\rm SPH}$ threshold at $z=2$ are included in the relations.  }
\label{fig:sub5}
\end{figure}

In our simulations, the SFRs and stellar masses of high redshift
galaxies are well correlated (see, e.g., figure 7 of \citealt{dave10}).
In the left panels of
Figure \ref{fig:sub5}, we directly investigate the relationship between the
$z=2$ SFRs of our simulated galaxies and the properties
of their parent halos, in the same format used previously for stellar
mass and $r$-band luminosity at $z=0$. To calculate the SFR of an SPH
galaxy, we sum over the SFRs of all its gas particles computed
according to the formula in \cite{katz96}. In contrast to $z=0$, where
there is a significant passive, non star forming population, only
$\sim0.1$ \% of galaxies have no star formation at $z=2$, and these
are excluded from the analysis. In the absence of dust extinction, the
instantaneous SFR should be a good indicator of rest-frame UV
luminosity, since the latter is dominated by the output of young,
short-lived stars. We caution, however, that dust extinction
corrections for LBG UV luminosities are typically factors of
several \citep[e.g.][]{steidel03}, and a scatter in extinction at
fixed SFR could add significant scatter to the relation between $M_H$
and UV luminosity. Since colours provide an indication of extinction,
the most effective strategy for observational analysis is probably to
apply colour based extinction corrections before subhalo abundance
matching, so that only the errors in the corrections add scatter.

Figure \ref{fig:sub5} shows that the relation between intrinsic SFR
and $M_H$ at $z=2$ is nearly as tight as the relation between $R$-band
luminosity and $M_H$ at $z=0$. We can therefore assign SFRs to SHAM galaxies 
via the same monotonic matching used previously for stellar mass and R-band
luminosity, with the results shown in the right panels.
The 68\% scatter of $|$log SFR$_A$/SFR$_R|$ is $\sigma_S$
= 0.12 and $\sigma_S$ = 0.16 for the SPHnw and SPHw simulations
respectively, compared to the $R$-band luminosity scatters of
$\sigma_L$ = 0.15 for both simulations at $z=0$. The median relations
for central and satellite galaxies are nearly the same, in both cases,
with offsets that are small compared to the intrinsic
scatter. However, in the no-wind simulation the outliers at low SFR
are preferentially satellites, a result of the gradual shutoff of gas
accretion after galaxies become satellites in larger
halos \citep{keres09,simha09}. The SFR$_A$/SFR$_R$ histogram is,
therefore, skewed towards overestimated SFRs for satellites, though
this remains a small effect. For the wind simulation, there are many
fewer galaxies above our 64 $m_{\rm SPH}$ stellar mass threshold, and
star formation rates at $M_H$ $<$ 10$^{12.4}$M$_{\odot}$ are
suppressed by the momentum driven outflows. In this simulation, there
are more satellite outliers at high SFR, and the (noisy) histogram of
SFR$_A$/SFR$_R$ for satellite galaxies shows an overall shift toward
underestimated SFR. For central galaxies, gas fractions in the SPHw
simulation are systematically higher than those in the SPHnw
simulation, and scatter in these central galaxy gas fractions
(contributing scatter in SFR) is the largest factor driving higher
$\sigma_S$ for the SPHw case.

Figure~\ref{fig:sub5} suggests that SHAM should be a reliable tool
for assigning SFRs to simulated halo populations at high redshift
(given an observed SFR distribution), or for inferring the halo
masses associated with observed Lyman-break galaxies.
We caution that our simulations do not produce a substantial
population of passive galaxies at these redshifts, either central
or satellite, and that physical mechanisms that strongly suppress
star formation at high stellar mass could reduce the
accuracy of SHAM assignment.


\section{SHAM in the DM Simulation}

So far we have applied subhalo abundance matching to the halo and
subhalo hosts of galaxies in the SPH simulations, where the condensed
baryons may improve the survival of dark matter substructures. We now
turn to SHAM as it is traditionally applied by using the DM
simulation, which starts from the same initial conditions but includes
no baryons. Independent SO halos above the resolution limit have
similar locations and masses in our SPHnw, SPHw and DM simulations. We
identify dark matter substructures within these SO halos using the
AdaptaHOP code as described earlier. While there is reasonable
agreement in the number and abundance of subhalos between the two SPH
simulations and the DM simulation, there are positional
differences. For each subhalo above our resolution threshold of 64
particles, we trace its high redshift progenitors up to the epoch when
it was an independent halo, i.e., up to \zsat.

\begin{figure}
\centerline{
\epsfxsize=5.5truein
\epsfbox{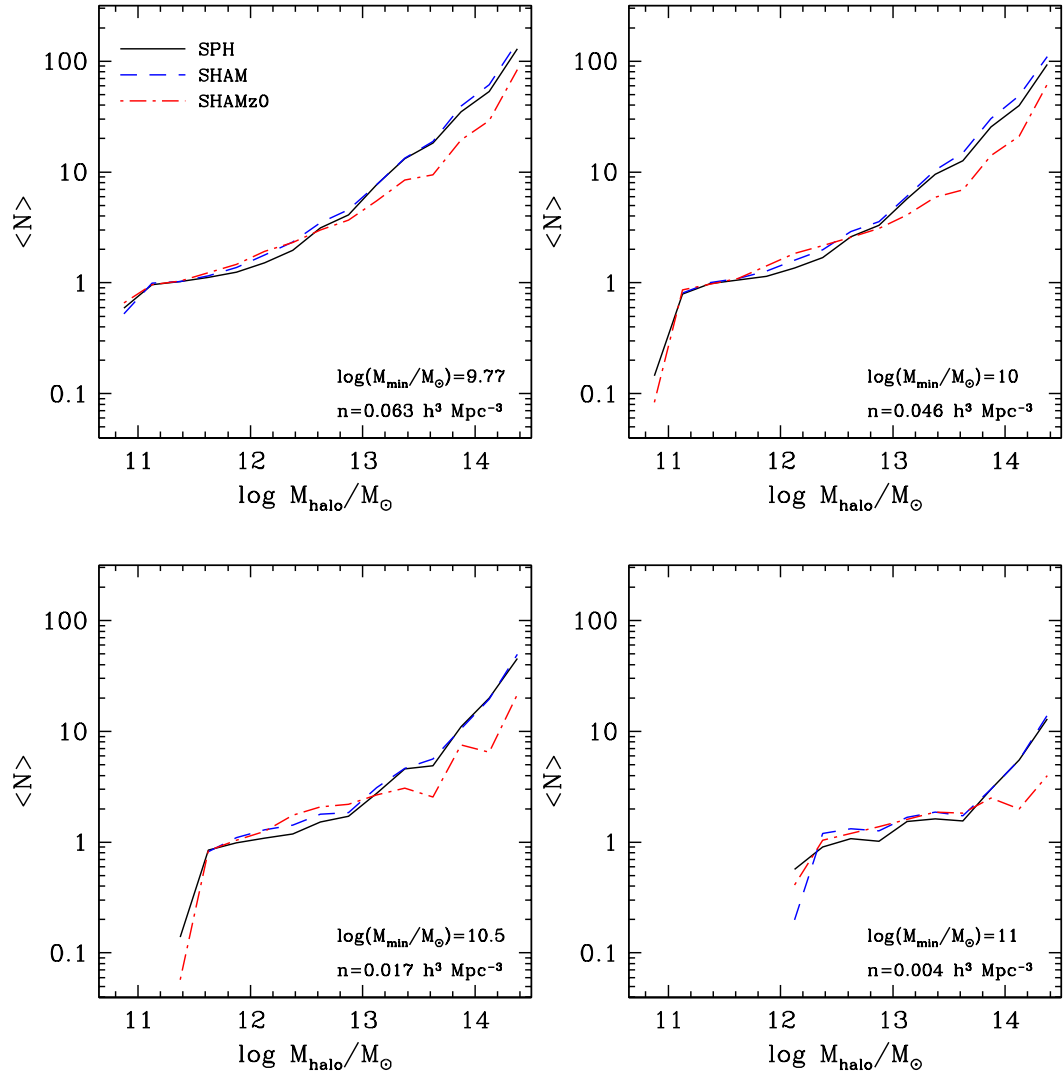}
}
\caption{
Mean number of galaxies per halo versus halo mass. The solid curve
represents the SPHnw simulation, while the dashed and dotted curves
are results from populating halos in our N-body simulation with
galaxies using subhalo abundance matching (SHAM) with subhalo masses
at \zsat and at $z=0$ (SHAMz0), respectively. Each panel stands for a
different galaxy stellar mass threshold, for which the corresponding
space density of galaxies is also indicated. Curves are truncated when
the mean occupation of halos in the next lower (0.1-dex) mass bin
falls below 0.03.  }
\label{fig:sub6}
\end{figure}

The solid curve in each panel of Figure \ref{fig:sub6} shows
$\langle$N(M)$\rangle$, the mean number of SPH galaxies per halo above
a given stellar mass threshold in each halo mass bin in the SPHnw
simulation. The four panels correspond to different stellar mass
thresholds, and the mean space density of galaxies above these
thresholds ranges from 0.004 to 0.06 $h^3$Mpc$^{-3}$. The dashed curve
shows the mean number of galaxies per halo when halos in the DM
simulation are populated with galaxies using SHAM as described in \S3,
with a monotonic relation between stellar mass and $M_H$ (equation
2). For comparison, the dotted curves show results of a model (SHAMz0)
in which we assume a monotonic relationship between galaxy stellar
mass and $z=0$ subhalo mass, rather than $z_{\rm sat}$ subhalo
mass. Note that the procedures of \cite{vale04} and \cite{weinberg08}
differ from SHAMz0 because they apply a global mass-loss correction to
subhalo masses, though they do not consider post-\zsat mass loss on an
object-by-object basis.

In low mass halos, where the satellite fraction is low, both the
SHAMz0 and SHAM model predictions for the average number of galaxies
per halo are in good agreement with the SPH simulation. At high halo
mass, however, SHAMz0 underpredicts the number of galaxies per halo
because it does not account for subhalo mass loss. The stellar masses
assigned to subhalos are too low, and galaxies that should be above a
stellar mass threshold instead fall below it. SHAM, on the other hand,
remains in good agreement with the SPH simulation over a wide range of
halo masses and galaxy stellar mass thresholds.

\begin{figure}
\centerline{
\epsfxsize=5.5truein
\epsfbox{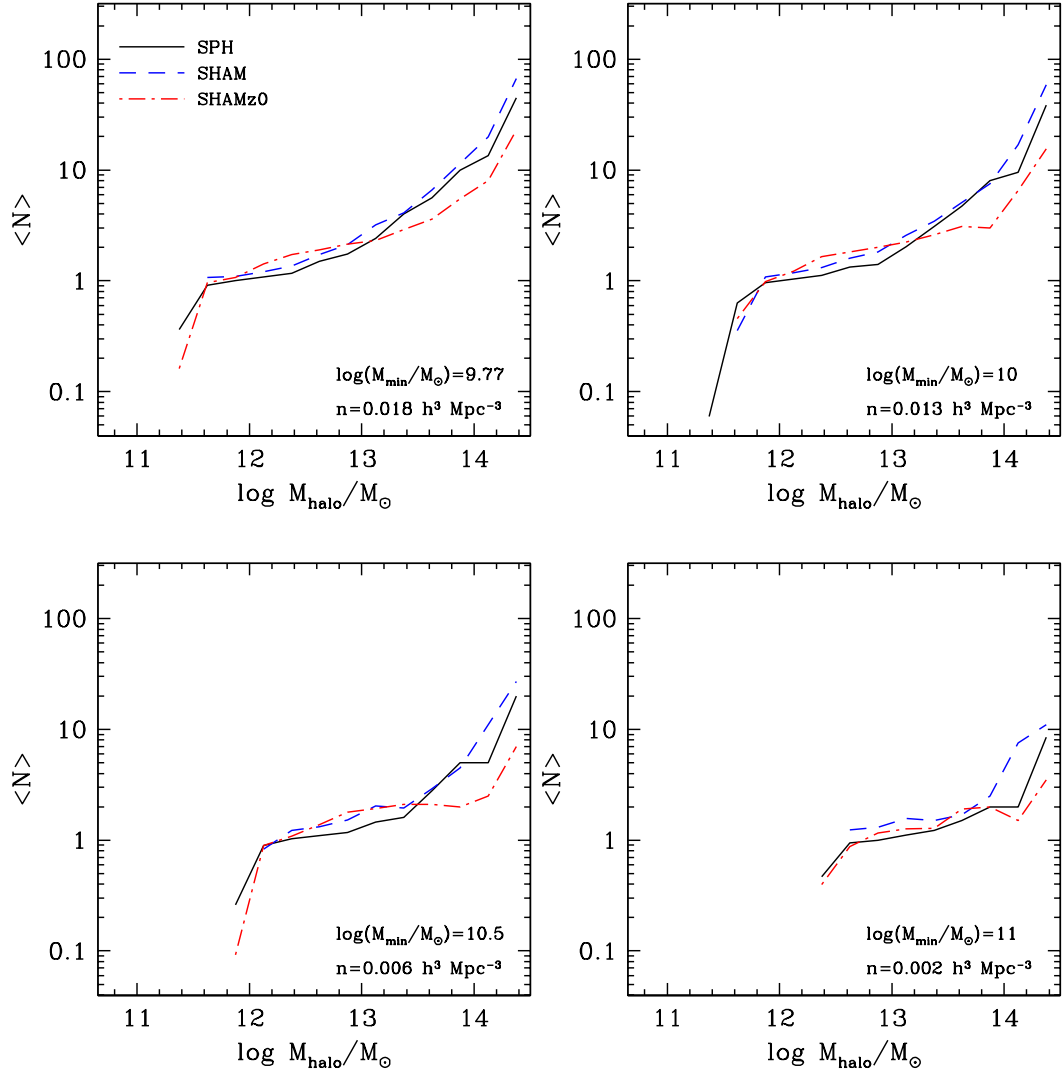}
}
\caption{
Mean number of galaxies per halo versus halo mass. This figure is
analogous to figure \ref{fig:sub6} but using the SPHw simulation,
which includes momentum driven winds.  }
\label{fig:sub7}
\end{figure}

Figure \ref{fig:sub7} is the analogue of Figure \ref{fig:sub6}, but
using the SPHw simulation instead of the SPHnw simulation. As winds
reduce stellar mass, the number density of galaxies above a given
stellar mass threshold is lower than in the SPHnw
simulation. Nonetheless, the trends for the SPHnw simulation discussed
above still hold. Abundance matching using the $z=0$ mass of subhalos
underpredicts the number of galaxies in massive halos. Although, SHAM
(using the z$_{\rm sat}$ mass of subhalos) provides a better fit to
the SPH galaxy sample, it overpredicts the number of galaxies in
massive halos to a larger degree than in the SPHnw simulation. The
one-halo term of the galaxy correlation function depends on the mean
pair number $\langle$$N(N-1)$$\rangle$. We have checked that the
standard prescription \citep{kravtsov04,zheng05} that satellite
galaxies follow Poisson statistics
($\langle$\Nsat(\Nsat-1)$\rangle$=$\langle$\Nsat$\rangle$$^2$
describes both simulations well, allowing computation of
$\langle$$N(N-1)$$\rangle$ from $\langle$$N(M)$$\rangle$.

\begin{figure}
\centerline{
\epsfxsize=5.5truein
\epsfbox{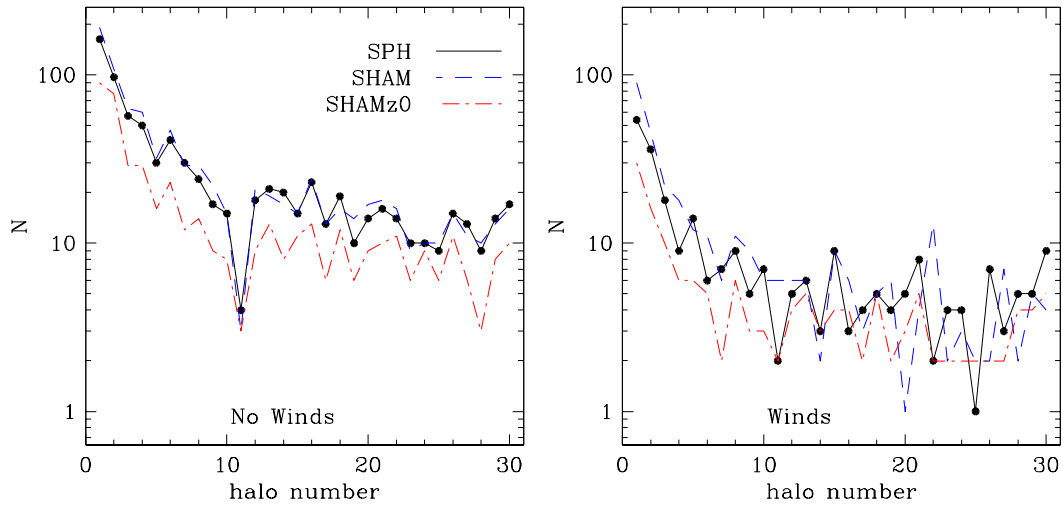}
}
\caption{
Halo occupations of the 30 most massive halos in the SPHnw (left) and
SPHw(right) simulations. Points connected by the solid curve represent
SPH galaxies, while the dashed and dotted curves represent SHAM and
SHAMz0 respectively. All galaxies above the resolution threshold are
included in this figure.  }
\label{fig:sub9}
\end{figure}

Figure \ref{fig:sub9} compares the halo occupations of SPH galaxies to
the SHAM and SHAMz0 populations in each of the 30 most massive halos
at $z=0$, for galaxies above the 64 $m_{\rm SPH}$ =
5.8$\times$10$^9$M$_{\odot}$ threshold. As already seen in
Figures \ref{fig:sub6} and \ref{fig:sub7}, SHAMz0 predicts too few
galaxies in massive halos in both simulations. In the SPHnw
simulation, the agreement between the number of SPH galaxies in each
halo and the number of galaxies assigned to the halo by SHAM is
remarkably good, indicating that SHAM is not just reproducing the
typical number of galaxies for a given halo mass, but is also
capturing the variation in galaxy number at a given halo mass. In the
SPHw simulation, however, SHAM more noticeably overpredicts the number
of galaxies in the most massive halos, and it does not track the
variation in galaxy number at a given halo mass as well as in the
SPHnw simulation.

\begin{figure}
\centerline{
\epsfxsize=5.5truein
\epsfbox{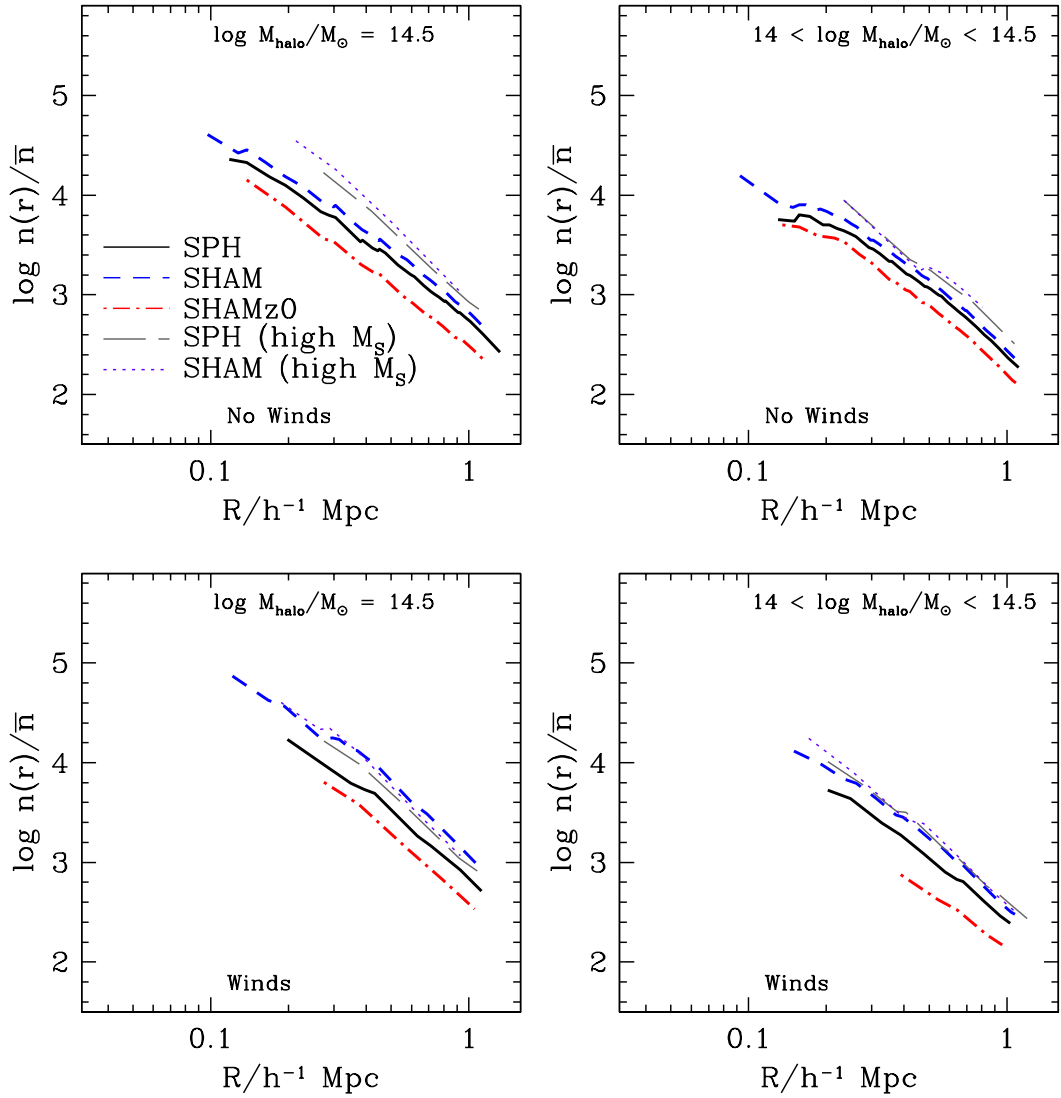}
}
\caption{
Radial number density profile of SPH galaxies (black solid curve) and
halos in the N-body simulation populated with galaxies using SHAM
(blue dashed curve) and SHAMz0 (red dot-dashed curve), for the mass
threshold of $M_S$ = 5.8 $\times$ 10$^9$ M$_{\odot}$. The top two
panels are for two halo mass bins in the SPHnw simulation while the
bottom two panels are for two halo mass bins in the SPHw
simulation. The curves stop when the only interior galaxy is the
central galaxy. Also shown are radial number density profiles of SPH
galaxies (gray solid curve) and SHAM (purple dashed curve) selected
halos using a higher threshold of $M_S$ = 7.3 $\times$10$^{10}$
M$_{\odot}$ in the no winds simulation and $M_S$ = 2.8
$\times$10$^{10}$ M$_{\odot}$ in the winds simulation.  }
\label{fig:sub8}
\end{figure}

Galaxy clustering depends on halo occupation statistics like those
shown in Figures 7-9, and on small scales it also depends on the
radial profile of satellites in massive halos. Figure \ref{fig:sub8}
compares the radial number density profile of SPH galaxies around the
central galaxy of the halo to the radial number density profile of
galaxies assigned to halos by SHAM and SHAMz0. The left panels
correspond to the single most massive halo in the simulations, with
$M_{\rm halo}$ = 4$\times$10$^{14}$M$_{\odot}$, while the right panels
correspond to the three halos with 10$^{14}$M$_{\odot}$ $<$ $M_{\rm
halo}$ $<$ 3$\times$10$^{14}$M$_{\odot}$. In all cases, the slopes of
the radial density profiles are similar for SPH galaxies and for
subhalos populated by SHAM and by SHAMz0, but the normalisations and
the inner truncations (indicated by where the curves stop) differ. The
normalisation differences correspond to the differences in halo
occupation at high $M_{\rm halo}$ seen in Figures \ref{fig:sub7}
and \ref{fig:sub9}. SHAMz0 predictions are always suppressed relative
to SPH galaxies because they neglect subhalo mass loss. Tidal
stripping is more severe for subhalos near the halo centre,
exacerbating this effect and causing truncation of the SHAMz0 profiles
at larger radii compared to SPH galaxies. \cite{weinberg08} found a
similar truncation effect even when including a global correction for
subhalo mass loss.

In contrast to SHAMz0, SHAM overpredicts the number of galaxies in
high mass halos, by a small factor in the SPHnw simulation, and by
0.1-0.3 dex in the SPHw simulation. This overprediction is a
consequence of neglecting stellar mass loss that occurs after \zsat in
the SPH simulation, which is not accounted for in the SHAM recipe. The
stellar mass loss is more severe for satellites close to the halo
centre, so in this case the SPH profiles truncate at larger radii than
the SHAM profiles.



The overprediction of satellite numbers by SHAM may seem surprising in
light of the good agreement between assigned and true stellar masses
seen in Figure 2, with distributions that peak at log $M_A$/$M_R$ = 0
and have only mild asymmetry. However, the relations in Figure 2 (and
the versions divided by halo mass in Figure 4) only include galaxies
that remain above the 64 $m_{\rm SPH}$ stellar mass threshold at
$z=0$. In high mass halos, a significant fraction of satellites have
dropped below the mass threshold by $z=0$ in the wind simulation, but
SHAM would place all of these galaxies on the $M_S$-$M_H$ relation
traced by the bulk of the satellites. In the no-wind simulation, the
fraction of satellites that suffer such severe mass loss is smaller.

We suspect that galaxies in the wind simulation are more fragile
because of their lower baryonic masses, and because they are typically
less concentrated than galaxies formed without winds; in both
respects, they more closely resemble observed galaxies. However,
satellites in the SPHw simulation that are disrupted are close to the
resolution threshold, and we cannot rule out the possibility that
their disruption is a numerical artifact. For comparison, in
Figure \ref{fig:sub8}, we also show the radial number density profile
of SPH galaxies and SHAM selected subhalos around the central galaxy
of the halo only for galaxies above a higher mass threshold of
7.3$\times$10$^{9}$M$_{\odot}$ in the no winds simulation and
2.8$\times$10$^{9}$M$_{\odot}$ in the winds simulation, corresponding
to a number density of 0.007 $h^3$ Mpc$^{-3}$. With the higher stellar
mass threshold, the differences in the normalisation and inner
truncation of the SPH and SHAM curves is substantially
reduced. Comparisons of our results with future higher resolution
simulations of a similar volume will help clarify the extent to which
our finite mass resolution affects our results.

\begin{figure}
\centerline{
\epsfxsize=5.5truein
\epsfbox{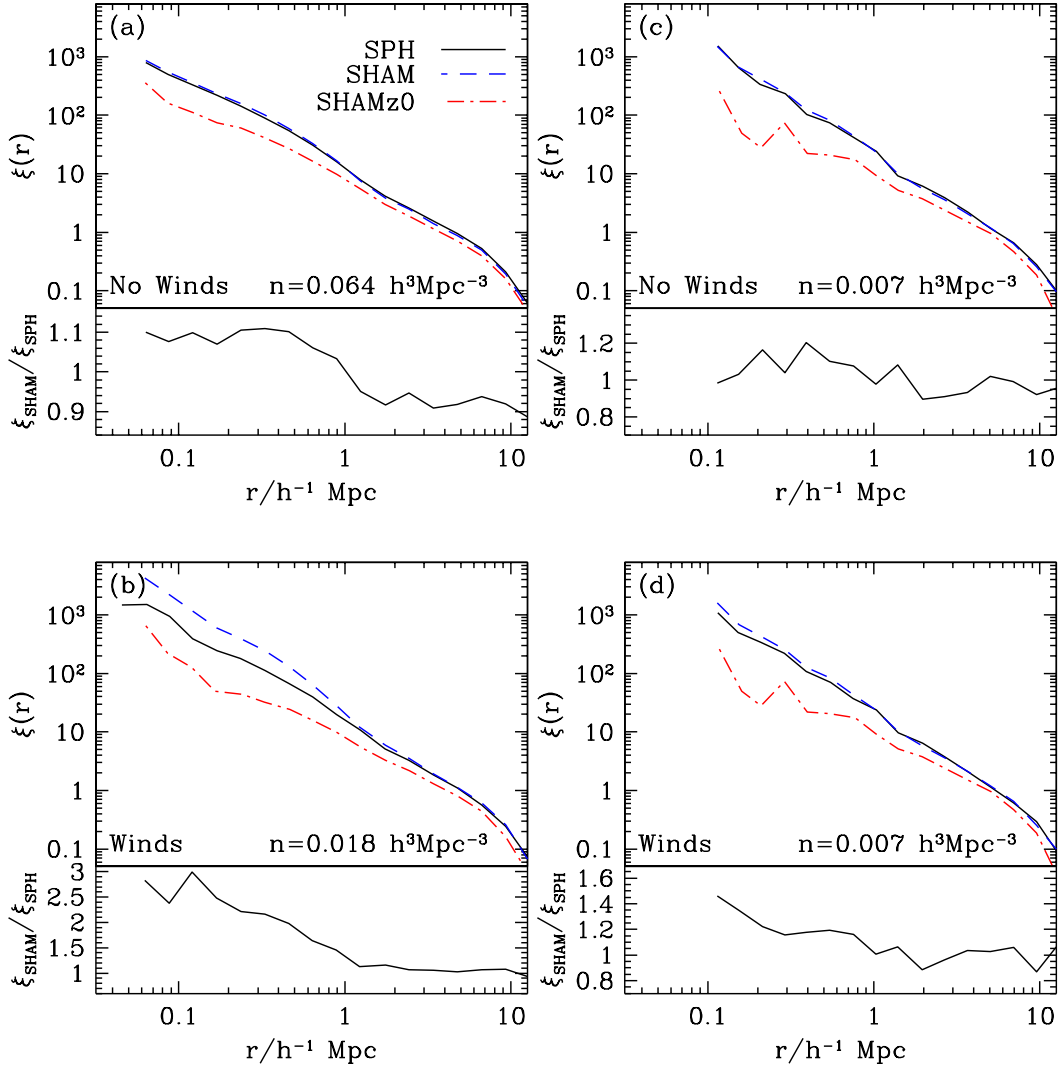}
}
\caption{
Two-point correlation function of SPH galaxies (solid curve) and halos
and subhalos in the N-body simulation populated using SHAM (dashed
curve) and SHAMz0 (dot-dashed curve). Panels (a) and (b) show all
galaxies above the stellar mass resolution threshold (5.8 $\times$
10$^9$ M$_{\odot}$) in the SPHnw (no winds) and SPHw (winds)
simulations respectively. Panels (c) and (d) show galaxies above a
10$^{12}$ M$_{\odot}$ halo mass threshold (at $z=0$ for central
galaxies or $z_{\rm sat}$ for satellite galaxies). Lower windows of
each panel show the ratio of the SHAM correlation function to the SPH
galaxy correlation function.  }
\label{fig:sub10}
\end{figure}

Panels (a) and (b) of Figure \ref{fig:sub10} show the two-point
correlation function of SPH galaxies (solid curve) and of halos and
subhalos selected by SHAM (dashed curve) and by SHAMz0 (dot-dashed
curve), in the SPHnw simulation and the SPHw simulation
respectively. All galaxies above our adopted stellar mass threshold of
64 $m_{\rm SPH}$ = 5.8$\times$10$^{9}$M$_{\odot}$ are included which
gives a space density of galaxies of 0.064 $h^3$ Mpc$^{-3}$ in the
SPHnw simulation and 0.018 $h^3$ Mpc$^{-3}$ in the SPHw
simulation. SHAMz0 underpredicts the correlation function in both
simulations, more severely for SPHw, as expected from its
underpopulation of massive halos (Figures 7-10). At scales r $\geq$
2h$^{-1}$Mpc, the ``two-halo'' regime of the correlation function, the
depression of $\xi(r)$ reflects a drop in the galaxy bias factor from
underweighting these highly biased halos. At scales $r$ $\leq$
1$h^{-1}$Mpc, where galaxy pairs within the same high mass halo make a
large contribution, the suppression of $\xi(r)$ is more severe.

For the SPHnw simulation, agreement between the SHAM and SPH-galaxy
correlation functions is excellent, as one would expect from the close
agreement of halo occupations and radial profiles seen in earlier
figures. The largest discrepancies are $\sim$10\% for both
2$h^{-1}$Mpc $<$ $r$ $\leq$ 10$h^{-1}$Mpc and $r$ $\leq$
2$h^{-1}$Mpc. For SPHw, the agreement in the 2-halo regime is still
very good, with a maximum discrepancy of $\sim$15\% for 2$h^{-1}$Mpc
$<$ $r$ $\leq$ 10$h^{-1}$Mpc. However, the overpopulation of high mass
halos leads to substantial overprediction of $\xi(r)$ in the 1-halo
regime, by up to a factor of $\sim$2.5. This discrepancy arises from
the severe stellar mass loss that affects a small but not negligible
fraction of satellites in high mass halos and is not captured by the
SHAM recipe. The impact on three-point or higher order correlation
functions, which more strongly weight the single-halo occupations at
small scales, would be more severe.

Panels (c) and (d) of Figure \ref{fig:sub10} are the analogues of
panels (a) and (b) respectively, but only show galaxies whose parent
halos are more massive than 10$^{12}$M$_{\odot}$ (at $z=0$ for central
galaxies and $z_{\rm sat}$ for satellite galaxies). This corresponds
to a space density of galaxies of 0.007 $h^3$ Mpc$^{-3}$ and a stellar
mass threshold of 7.3$\times$10$^{10}$M$_{\odot}$ in the no winds
simulation and 2.8$\times$10$^{10}$M$_{\odot}$ in the winds
simulation. For the SPHnw simulation, the trends discussed above still
hold, agreement between the SHAM and SPH-galaxy correlation functions
is better than $\sim$18\% on all scales. For the SPHw simulation, in
contrast with panel (b), the agreement between the SHAM predicted and
SPH galaxy correlation functions is better than $\sim$20\% on all
scales except $r$ $\leq$ 0.3$h^{-1}$Mpc, where the discrepancy rises
to $\sim$50\%. Compared to galaxies hosted by low mass subhalos, a
substantially smaller fraction of galaxies hosted by more massive
subhalos undergo severe stellar mass loss. Furthermore, for subhalos
with $M_H$ (at \zsat) $\geq$ 10$^{12}$M$_{\odot}$, the fraction of
satellites that survive up to $z=0$ is only marginally lower in the
SPHw simulation compared to the SPHnw simulation.

\section{Conclusions}

Ever since the identification of substructures in N-body
simulations \citep[e.g.][]{ghinga98,klypin99,moore99,springel01},
there have been efforts to associate them with
galaxies \citep[e.g.][]{colin99,kravtsov04,weinberg08}. Subhalo
abundance matching (SHAM) has had impressive empirical success,
reproducing observed galaxy clustering over a wide range of luminosity
and redshift and correctly diagnosing (via cluster mass-to-light
ratios) the overestimated matter clustering amplitude
($\sigma_8\approx0.9$ vs. $\sigma_8\approx0.8$) in WMAP1-era
cosmological models \citep{conroy06,vale06,guo10,moster10}. Our
investigation provides the first full-scale test of SHAM against
hydrodynamic cosmological simulations, where the correct
identification between galaxies and subhalos is known a priori,
including both a simulation with minimal feedback (SPHnw) and a
simulation with momentum-driven winds that better reproduces the
observed galaxy stellar mass function \citep{oppenheimer10}. This
investigation yields physical insight into the galaxy formation
process in these simulations, and it demonstrates the strengths and
potential limitations of SHAM as a tool for interpreting observed
galaxy clustering, significantly extending earlier studies
by \cite{nagai05} and \cite{weinberg08}.

When we consider galaxies above our adopted stellar mass threshold at
$z=0$, $M_S$ $\ge$ 64 $m_{\rm SPH}$ = 5.8$\times$10$^9$ M$_{\odot}$,
we find a tight correlation between galaxy stellar mass and the mass
of the parent halo or subhalo. For central galaxies, we consider the
full mass of the spherical overdensity at $z=0$, while for satellite
galaxies we use the mass of the parent halo just before the
epoch \zsat when it first becomes a satellite. Importantly, the median
relation and scatter between $M_S$ and $M_H$ are similar for central
galaxies and satellite galaxies, in both simulations, and the outlier
fraction for satellite galaxies is only modestly higher, mainly
because of stellar mass loss in some satellites after \zsat. As a
result, SHAM assignment of stellar masses is remarkably effective in
both simulations. The 68\% scatter in $R$ = log $M_A$/$M_R$, where
$M_A$ is the assigned stellar mass and $M_R$ the real (simulation)
stellar mass, is $\sigma_M$ = 0.09 dex for SPHnw and 0.11 dex for
SPHw. The distribution of $R$ is only mildly asymmetric, with a small
extended tail of outliers. Similar results, with slightly increased
scatter, hold at $z=$ 0.5, 1 and 2. We find no clear correlation
between residuals from the $M_S$-$M_H$ relation and the satellite
epoch at \zsat or the parent mass of the $z=0$ halo, so our tests do
not suggest a way to further tighten SHAM by considering additional
properties.

Using $R$-band luminosity in place of stellar mass yields similar
results, but with a larger population of outliers among satellites
caused by their systematically older stellar populations and thus
higher $M_S/L$ ratios. SHAM should, therefore, be applied to stellar
masses (estimated from luminosity and colour or spectral energy
distribution) when possible, or to luminosity in redder bands that
more faithfully trace stellar mass. At $z=2$, instantaneous star
formation rates, which should be a good proxy for observed-frame
optical luminosities, are well correlated with $M_H$, with similar
correlations for central and satellite galaxies. SHAM assignment of
star formation rates at this redshift is quite effective with a
scatter in log SFR$_A$/SFR$_R$ of 0.12 dex in SPHnw and 0.16 dex in
SPHw. This result reinforces empirical evidence, based on galaxy
clustering data, that SHAM is an effective tool for modeling
Lyman-break galaxies at high redshift.

SHAM is traditionally applied to N-body rather than SPH simulations
(or to analytic descriptions calibrated on N-body). To test this
standard form of SHAM, we have applied it to the AdaptaHOP subhalo
population of a pure dark matter (DM) simulation started from the same
initial conditions as the SPH simulations. Because subhalo positions
within parent halos shift between SPH and DM \citep{weinberg08}, we
have focused our comparison on halo occupation statistics and radial
profiles, which together determine many properties of observable
galaxy clustering, and on the real space two-point correlation
function.

Using the galaxy stellar mass function of the SPHnw or SPHw simulation
as input, SHAM (applied to the DM simulation) does quite well in
reproducing the corresponding mean halo occupation, $\langle$$N(M_{\rm
halo}$)$\rangle$, and the slope of the galaxy radial profile in high
mass halos. In SPHnw, but not SPHw, SHAM traces the individual
halo-to-halo variations in galaxy number at similar halo mass. Use of
subhalo masses at \zsat rather than $z=0$ makes a critical difference;
if we use $z=0$ subhalo masses, the galaxy occupation in high mass
halos is systematically depressed because subhalos lose mass by tidal
stripping after becoming satellites. Traditional SHAM (using \zsat
masses), by contrast, tends to overpredict the galaxy numbers in high
mass halos. We trace this discrepancy to the small but not negligible
population of satellite galaxies that suffer severe stellar mass loss
in the SPH simulations, so that they move from above our resolution
threshold at \zsat to below it at $z=0$. The subhalos themselves
remain identifiable in the DM simulation, and SHAM populates them with
galaxies that lie on the main $M_S$-$M_H$ relation. The effect is more
significant in the SPHw simulation, whose galaxies are apparently more
vulnerable to severe mass loss because of their shallower baryonic
potential wells. The mean occupation of high mass halos in SPHw is
overpredicted by 0.1-0.3 dex, while in SPHnw the offset is less than
0.1 dex. For SPHnw, the SHAM-predicted correlation function agrees
with that of SPH galaxies to better than 10\% at all scales 0.05
$h^{-1}$Mpc $<$ $r$ $<$ 10 $h^{-1}$Mpc. For SPHw, SHAM predicts the
correlation function to 15\% at $r$ $>$ 2 $h^{-1}$Mpc but it
overpredicts by a factor of $\sim$2.5 at $r$ $<$ 0.5 $h^{-1}$Mpc
because of the overpopulation of massive halos. However, if the
stellar mass threshold is increased to 2.7$\times$10$^{10}$
M$_{\odot}$, then the SHAM predicted two-point correlation function
agrees with that of SPH galaxies to better than 20\% on all scales
except $r$ $<$ 0.3 $h^{-1}$Mpc, where the discrepancy widens to
50\%. This improvement is achieved because galaxies that suffer severe
mass loss are typically low stellar mass galaxies in low mass subhalos
($M_H$ $\leq$ $10^{12}$ M$_{\odot}$). While we would expect low mass
galaxies in low mass subhalos to be more susceptible to severe mass
loss, further investigation with higher resolution simulations is
required to understand the extent to which this is a numerical
artifact caused by proximity of the mass to the threshold above which
we are reliably able to resolve galaxies. For galaxies with luminosity
L$>$L$_*$, our results suggest that SHAM will modestly (by $\sim$
10-30 \%) overpredict the two-point correlation function on small
scales and yield accurate clustering predictions on larger scales.

In both of our SPH simulations, halo mass (at $z=0$ for central
galaxies or \zsat for satellite galaxies) is the primary determinant
of galaxy stellar mass, luminosity, and (at high redshift) star
formation rate. The most significant secondary factor, and thus the
most significant limitation on the accuracy of SHAM, is the small
fraction of satellite galaxies that suffer severe stellar mass loss
even though their host subhalos survive. The sensitivity to feedback,
indicated by the difference between our SPHnw and SPHw simulations,
motivates further investigation of galaxy mass loss in high mass halos
for a wider range of feedback prescriptions.

For galaxies with $L$ $\geq$ $L_*$, the SPHw simulation predicts
excessive galaxy masses and excessive late-time star formation,
indicating the need for an additional physical mechanism such as AGN
feedback. The accuracy of SHAM for high luminosity galaxies will
depend on how tightly correlated such feedback is with halo
mass. Overall, however, the strong role of dark matter in governing
galaxy formation makes subhalo abundance matching a powerful technique
for making realistic artificial galaxy catalogues and for interpreting
the observed distribution of galaxies.

\section*{ACKNOWLEDGEMENTS}
We thank Stephane Colombi, Charlie Conroy, and Andrey Kravtsov for
useful discussions on these topics. We thank Stephane Colombi for
providing the AdaptaHOP code and for assistance in using it. We thank
Charlie Conroy for providing the stellar population code used to
compute galaxy luminosities. This work was supported by NSF grant
AST0707985 and NASA ATP grant NNX1OAJ956.

\clearpage
\onecolumn


\clearpage
\bibliographystyle{mn2e}

\end{document}